\documentclass[aps,preprint,showpacs,amsmath,amssymb]{revtex4}

\usepackage{graphicx,amsmath,amssymb}

\newcommand{\be}{\begin{equation}}
\newcommand{\ee}{\end{equation}}

\newcommand{\vecp}{{\mathbf p}}
\newcommand{\vecq}{{\mathbf q}}

\newcommand{\vxi}{{\mathbf \xi}}
\newcommand{\x}{{\mathbf x}}

\newcommand{\vl}{{\mathbf l}}	

\newcommand{\J}{{\mathbf J}}

\newcommand{\opR}{{\hat{R}}}

\newcommand{\oprho}{{\hat{\rho}}}

\newcommand{\opL}{\hat{L}}

\newcommand{\opx}{\hat{\mathbf x}}

\newcommand{\M}{{\mathbf M}}

\newcommand{\der}{\partial}

\newcommand{\vct}[1]{\ensuremath\mbox{\boldmath$ #1 $}}
\newcommand{\Vxi}{\vct \xi}

\begin{document}
\title{Vestiges of quantum oscillations in the open evolution of semiclassical states}

\author{A. M. Ozorio de Almeida\footnote{Corresponding author}}
\address{Centro Brasileiro de Pesquisas Fisicas; \\
Rua Xavier Sigaud 150, 22290-180, Rio de Janeiro, RJ,
Brazil.\\ozorio@cbpf.br}

\begin{abstract}

A single wave component of a quantum particle can in principle be detected by the way 
that it interferes with itself, that is, through the local wave function correlation.
The interpretation as the expectation of a local translation operator allows this
measure of quantum wavyness to be followed through the process of decoherence in
open quantum systems. This is here assumed to be Markovian, determined by 
Lindblad operators that are linear in position and momentum.
The limitation of small averaging windows and even smaller correlation lengths
simplifies the semiclassical theory for the evolving local correlation. Its spectrum
has a peak for each classical momentum, subjected to Gaussian broadening with decoherence.
These spectral lines can be clearly resolved even after the Wigner function has become
positive: The correlations located far from caustics seem to be the last vestige of
quantum oscilations.

\end{abstract}

\maketitle

\section{Introduction}

Waves and the dynamics of particles were separately studied in classical physics before the birth
of quantum mechanics. One of the strange novelties that then came into being is the imputation
of a certain wave nature for a single particle, once the unity of classical phase space is
split into complementary descriptions, either in positions or momenta: A momentum eigenstate
is represented by a $\delta$-function in its own representation, but it is a plane wave
in the position representation and vice versa. In the full quantum picture in phase space
provided by the Wigner function \cite{Wigner}, these simple states appear as
just a $\delta$-function along the corresponding plane, so that the implicit wavyness 
can only be elicited by a (symmetrized) Fourier transform into either the position 
or momentum coordinates.

The standard way to examine the properties of waves is through interference. The linear superposition
of position or momentum eigenstates leads to structures that are evident in the intensities 
of their wavy representations and in the Wigner function itself. 
Furthermore, the mere dynamical bending of the supporting plane into a {\it Lagrangian surface} 
already provokes interferences in either of these representations.
Interferences are considered to be {\it quantum resources} in the language of quantum information,
but they are very sensitive to dephasing or decoherence in contrast to the intrinsic measure
of {\it wavyness} for a single wave: A correlation matches the wave with its own translation, instead
of another wave. Thus, even a very small correlation length, of the order of its tiny quantum wavelength,
already detects the wavyness of a momentum eigenstate or of a single branch of a Lagrangian surface
that supports a more general state. It will then be seen that these are the last vestiges 
of quantumness that survive the process of decoherence.

Decoherence results from the contact of a quantum system with an uncontroled environment,
so that an initial pure state evolves into a mixture \cite{Giulini}. Quantum interference within
the initial state is quickly erased in this process of emergent
classicality, but the point at which the last vestige of wavyness is lost 
by the evolving density operator should also be enquired. Does this event occur
globaly, or is there an intermediate regime in which some robust type 
of local quantum oscilation can still be detected through the prevailing classicality?

The assumption of Markovian evolution, governed by a master equation, 
in which all the {\it Lindblad operators} \cite{Lindblad} responsible for the nonunitarity of the evolution 
are linear functions of the phase space variables,
narrows down the investigation while maintaining considerable relevance.
If, furthermore, the driving Hamiltonian is quadratic, whether harmonic or not, 
it is shown that the Wigner function describing the state becomes positive
at a definite time that does not depend on the initial state, providing
that it was not positive already \cite{BroAlm04}. But is this really the ultimate criterium 
for complete classicality? Is it not possible to evince the existence of local
oscilations even if they are not strong enough to push the Wigner function
into negative values?

The generalization of the exact results in \cite{BroAlm04}
to arbitrary Hamiltonians was carried out in \cite{AlmRiBro,BroAlm10,AlmBro11} 
as a semiclassical (SC) approximation, assuming that 
the initial state also has a SC form. In other words, the state should
correspond classically to a Lagrangian surface in phase space (typically a torus)
\cite{Arnold}, such as a Fock state in one or more dimensions. Then,
unlike the case of quadratic evolution, the final surface develops many new branches
as described by \cite{BerBal79} that contribute SC oscilatory terms to any given choice
of representation, e.g. position, momentum, or the standard phase space representations,
that is, the Wigner function and its Fourier transform, known as
the {\it chord function}, in the notation of \cite{Report}, or the {\it quantum charateristic function}.
The latter is also a complete representation of a quantum state,
while providing a complementary description: All traces of classical structures
are compressed into a small neighbourhood of the chord origin, so that large scales of
the chord function illuminate purely quantum nonlocal correlations.
Thus, in contrast to the Wigner function and the Husimi function \cite{Husimi,Takashi} 
that are traditional rivals for eliciting classical structures in quantum states, 
the chord function is a priviledged probe for delicate quantum features, 
such as coherence loss in open systems studied in \cite{AlmRiBro,BroAlm10,AlmBro11}.  

A pair of points on the Lagrangian surface defines a SC-relevant chord.
Decoherence quenches its contribution to the chord function by the exponential
of a negative functional of this chord's history. It is only in the quadratic case
that all chords have the same {\it decoherence functional}, leading to the Wigner
function as a simple convolution. But the Fourier transform into the Wigner function
depends on the entire chord function, even though all long chords are quickly quenched.
Thus, the key question is whether the surviving region of the chord function itself  
stores unequivocable quantum information near the origin for appreciable times.

It turns out that {\it local wavefunction correlaltions} (LWC) are the appropriate measure 
of residual wavyness. They were proposed by Berry \cite{Berry77b} 
in the context of quantum ergodicity \cite{Voros76, Shnirelman, Verdiere, Zelditch}
and here they are identified with the expectation of a {\it local translation operator},
which allows their generalization for mixed states. 
The LWC have also been found to have experimental relevance (see \cite{Jalabert} 
and references therein). Actually, the states that are prepared for an experiment
are generally simple eigenstates of integrable Hamiltonians, or their superpositions,
so that the aim here is to determine the effect of decoherence on the LWC 
of integrable eigenstates, which were also analyzed in \cite{Berry77b}, rather
than the correlations of {\it ergodic states} that have received most attention so far.

The fact that the LWC are entirely determined by the small chord
neighbourhood of the chord function \cite{ZamZapOA} heralds their suitability 
for the present study. It is shown in the Appendix that the complete set of LWC have the
same information content as the Husimi function if the averaging distances are the widths
of a coherent state. Thus a pure state is in principle completely defined by its LWC.

A special bonus of thus being able to restrict our study to small chords
is that then the general nonlinearity of the corresponding classical motion is approximately
subsumed within the time dependence of the quadratic expansion of the full Hamiltonian 
around each trajectory. The consequent linearized approximation vastly simplifies 
the full SC theory in double phase space that was developed in our previous work
\cite{AlmRiBro,BroAlm10,AlmBro11}, both for the unitary evolution and the decoherence functional.

The general common framework of both the Weyl-Wigner representation and the chord representation 
is reviewed in the following section before discussing the {\it classical basis} of the latter 
in the limit of small chords. Section 3 then recalls some basic aspects of the SC approximation
to the nonunitary evolution of the chord function within a Markovian framework. 
Section 4 presents the local translation operators and the LWC, together with the simple example 
for coherent states. One is then in a position to analyze the decoherence of the LWC in terms of the linearized motion of the classical trajectories of the relevant phase space points in section 5. 
The scenario for the final loss of any identifiable wavyness is then discussed in section 6. 
An Appendix bridges the Husimi function to the full set of LWC with averaging widths 
chosen of the order of $\hbar^{1/2}$.

\section{The short chord limit}

Let us recall that $\mathbb{R}^{2N}$ stands for a $(2N)$-dimensional classical phase space, 
$\{\x=(\vecp, \vecq)\}$ with its {\it skew product}, 
\begin{equation}
 \x\wedge \x'=\sum_{l=1}^N
(p_l q'_l - q_l p'_l)= \J\> \x\cdot \x', 
\label{squew} 
\end{equation} 
which also defines the skew symplectic matrix $\J$. 
We shall here use a distinct notation for the {\it centre} of a pair of points,
$\x = (\x^+ + \x^-)/2$, whereas the {\it chords},
$\Vxi = (\Vxi_{\vecp},\Vxi_{\vecq}) = \x^+ - \x^-$, are the conjugate variables
to the {\it centres} $\x$ and correspond to tangent
vectors in phase space, as in the scheme for a Legendre transform.
Each of these chords lables a uniform translation of phase space
points $\x_0 \in \bf{R}^{2N}$ by the vector $\Vxi \in \bf{R}^{2N}$,
that is: $\x_0 \mapsto \x_0 + \Vxi$. Likewise, each centre, $\x$,
labels a reflection of phase space $\bf{R}^{2N}$ through the point $\x$, 
that is $\x_0 \mapsto 2\x - \x_0$.

Corresponding to the classical translations, one defines {\it translation operators},
\begin{equation}
\hat{T}_{\Vxi} = \exp\left\{\frac{i}{\hbar}\,\Vxi\wedge \hat{\x}\right\} 
= \int d \vecq \left| \vecq + \frac{\Vxi_{\vecq}}{2} 
\right\rangle \left\langle \vecq - \frac{\Vxi_{\vecq}}{2} \right|
                e^{{i \Vxi_{\vecp} \cdot \vecq}/ \hbar}   ,
\label{transop}								
\end{equation}
also known as displacement operators, or Heisenberg operators.
The chord representation of an operator $\hat{A}$ on the Hilbert
space ${L}^2(\bf{R}^N)$ is defined as a linear (continuous) superposition of translation operators.
In this way,
\begin{equation}
\label{conC} 
\hat{A} = \frac{1}{(2\pi\hbar)^N}\int d\Vxi 
\tilde{A}(\Vxi) \hat{T}_{\Vxi} \
\end{equation}
and the expansion coefficient, a function on $\bf{R}^{2N}$, is the
{\it chord symbol} of the operator $\hat{A}$:
\begin{equation}
\label{covC} 
\tilde{A}(\Vxi) = {\rm tr} \left[\hat{T}_{-\Vxi}\hat{A}\right]  .
\end{equation}

The Fourier transform of the translation operators defines the
{\it reflection operators},
\begin{equation}
 2^N\hat{R}_\x = \frac{1}{(2\pi\hbar)^N}\int d\Vxi 
\exp\left\{\frac{i}{\hbar}\,\x\wedge\Vxi\right\}  \hat{T}_{\Vxi}  
= \int d \vxi_q 
      \left| \mathbf q + \frac{\Vxi_q}{2} \right\rangle 
\left\langle \mathbf q - \frac{\Vxi_q}{2} \right|
       e^{i \mathbf p \cdot \Vxi_q / \hbar},
\label{refl}
\end{equation}
such that each of these corresponds classically to a reflection of phase space
$\bf{R}^{2N}$ through the point $\x$.
The same operator $\hat{A}$ can then be decomposed into a
linear superposition of reflection operators
\begin{equation}
\label{conW} 
\hat{A} = 2^N\int \frac{d\x}{(2\pi\hbar)^N} A(\x) 
\hat{R}_{\x} ,
\end{equation}
thus defining the {\it centre symbol or Weyl symbol} 
of operator $\hat{A}$ in the manner of Grossmann \cite{Grossmann}:
\begin{equation}
\label{covW} 
A(\x) = 2^N{\rm tr}\left[\hat{R}_{\x}\hat{A}\right].
\end{equation}
In the case of the density operator, it is convenient to use another normalization for 
the {\it Wigner function} \cite{Wigner,Royer} and the {\it chord function} as 
\begin{equation}
W(\x) = \frac{2^N}{(2\pi\hbar)^N}{\rm tr}\left[\hat{R}_{\x}\hat{\rho}\right]
~~ {\rm and}~~
\chi(\Vxi) = \frac{1}{(2\pi\hbar)^N}{\rm tr} (\hat{T}_{-\Vxi}\hat{\rho}). 
\label{Wtr} 
\end{equation} 

The centre and chord symbols are always related
by a Fourier transform:
\begin{equation}
 \tilde{A}(\Vxi) = \frac{1}{(2\pi\hbar)^N}\int d\x 
 \exp\left\{\frac{i}{\hbar}\,\x\wedge \Vxi\right\}
A(\x)
 \ ,
\label{FWS}
\end{equation}
\begin{equation}
 A(\x) = \frac{1}{(2\pi\hbar)^N}\int d\Vxi 
\exp\left\{\frac{i}{\hbar}\,\Vxi\wedge \x\right\} 
\tilde{A}(\Vxi).
\label{FCS}
\end{equation}
In particular, one obtains the reciprocal representations of the reflection operator and 
the translation operator as
\begin{equation} 
2^N\tilde{R}_{\x}(\Vxi)= \exp \left\{\frac{i}\hbar \,\x \wedge \Vxi\right\} {\rm ~or~}
T_{\Vxi}(\x) = \exp \left\{-{\frac{i}\hbar}\, \x \wedge \Vxi\right\}. 
\label{planewave} 
\end{equation}
These expressions are ideally suited for use in SC approximations. The direct representations are 
\begin{equation} 
2^N\tilde{R}_{\x}(\x')= \delta(\x'-\x) {\rm~or~}
T_{\Vxi}(\Vxi') = \delta(\Vxi'-\Vxi).  
\label{directrep}
\end{equation}

The basis for the following developments is the curious form of expressing the Fourier relation 
between the Wigner function and the chord function as
\begin{equation}
 \chi(\Vxi) = \frac{1}{(2\pi\hbar)^N}\int d\x ~W(\x)~ 2^N\tilde{R}_{\x}(\Vxi)\ .
\label{chi-R}
\end{equation}
The key point is that the substitution of the full oscillatory quantum Wigner, $W(\x)$ function
in this formula by its corresponding classical Liouville distribution, $W_c(\x)$, can be 
a good approximation for the true chord function in a neighbourhood of the origin.
Indeed, all polynomial moments, i.e. expectations of positions and momenta are just
multiple derivatives of $\chi(\Vxi)$ at the origin, so that whenever the averages for these variables
are approximately classical, so will be the short chord limit of the chord function. 

The nontrivial nature of this {\it short chord approximation} is manifest upon considering it 
as the chord symbol of the {\it classical density operator},
\begin{equation}
 \oprho_c = \int d\x 
~W_c(\x)~ 2^N\opR_{\x}\ .
\label{rho_c}
\end{equation}
The centre symbol for this expression, using \ref{directrep}, merely returns the original
Liouville distrbution, $W_c(\x)$. In contrast, the 'classical' approximation of \ref{chi-R}
is a sum over plane waves instead of $\delta$-functions.
In the case of a pure integrable state that can be described by a generalization of the WKB
approximation, so that it corresponds to a Lagrangian surface, this simple form of the 
short chord approximation does hold \cite{ZamOA10}. Furthermore, it can be extrapolated
to provide essentially nonclassical information concerning the state 
in the form of {\it blind spots}. These isolated zereos of the chord function 
are too sensitive for our present purposes, since they disappear immediately with the onset of decoherence. 
However, it is shown in \cite{ZamZapOA}
that local correlations of the wave function are also predicted by the short chord approximation.
These will be shown in the next section to be fairly robust with respect to decoherence.
The discussion of {\it quantum ergodicity} in \cite{ZamZapOA} extrapolates from the WKB
type of states and provides evidence that the short chord approximation holds
even in the absence of integrability. Here we will assume that it is quite general;
for instance, that it can also be assumed for an evolving coherent state as, say,
the corresponding Liouville density wraps around the unstable manifold of an unstable fixed point.

It should always be borne in mind that no information is assumed concerning 
the chord function outside a small neighbourhood of the origin, so that one is forbidden to take
its Fourier transform to what would be merely the starting point: $W_c(\x)$. This initial classical
distribution is just the 'scafolding' for the short chord approximation, not the
'building' itself!

\section{Markovian evolution of short chords}

The Lindblad master equation \cite{Lindblad} describes the general evolution for
markovian open systems under the weakest possible constraint. Given
the internal Hamiltonian, $\hat{H}$, and the Lindblad operators,
$\hat{L_k}$, which account for the action of the random
environment, the evolution of the density operator may be reduced
to the canonical form,
\begin{equation}
\frac{\partial \hat{\rho}}{\partial t} = -\frac{i}{\hbar}[\hat{H},
\hat{\rho}] + \frac{1}{\hbar}\sum_k
(\hat{L_k}\hat{\rho}\hat{L_k}^{\dag} -
\frac{1}{2}\hat{L_k}^{\dag}\hat{L_k}\hat{\rho} - \frac{1}{2}
\hat{\rho}\hat{L_k}^{\dag}\hat{L_k} ), \label{Lindblad}
\end{equation}
so that, in the absence of the environment ($\hat{L_k}=0$), the
motion is governed by the Liouville-Von Neumann equation
appropriate for unitary evolution. Henceforth, we shall ommit the sum and the index, $k$, since
the results depend linearly on them.

In the chord space, by using product rules for the product of operators \cite{Report}, 
the Lindblad equation is represented by a partial differential equation. 
In the case where the Lindblad operators are linear functions of $\hat\vecp$ and $\hat\vecq$, that is, 
\begin{equation}
\opL = \vct l'\cdot\opx + i\vct l''\cdot\opx,  
\label{linlin}
\end{equation}
with $\vct l'$ and $\vct l''$ real vectors, the master equation (for a single $k$-component) 
can be written as \cite{BroAlm10}
\begin{eqnarray}
\frac{\der \chi}{\der t} (\vct \xi,t) = 
&& -\frac{i}{\hbar} {\mathcal N}\int \Bigl[H(\x+\frac{1}{2}\vct \xi,t)-H(\x-\frac{1}{2}\vct \xi,t)\Bigr]
~\exp{ \left(\frac{i}{\hbar}(\vct \xi'-\vct \xi)\wedge \x \right)}
~\chi(\vct \xi',t)
~d\vct \xi'~d\x \nonumber \\
&& -\gamma ~\vct \xi\cdot \frac{\der \chi}{\der \vct\xi}(\vct \xi,t) 
- \frac{1}{2\hbar}~\Bigl[(\vl'\cdot\vct \xi)^2 + (\vl''\cdot\vct \xi)^2 \Bigr] 
~\chi(\vct \xi,t). 
\label{dynchi}
\end{eqnarray}
The {\it dissipation coefficient}, 
\begin{equation}
\gamma = \vl''\wedge\vl',
\label{gamma}
\end{equation}
is null for a Hermitian Lindblad operator ($\vl''=\vct 0$) and we then have a purely diffusive case. 
$H$ is the Weyl representation of the Hamiltonian of the isolated system 
and coincides with the corresponding classical Hamiltonian, 
up to semiclassically small corrections due to the of non-commutativity of $\hat\vecp$ and $\hat\vecq$.
By defining the variables  $\vct y = \mathbf J\vct\xi = (-\Vxi_{\vecq},\Vxi_{\vecp})$, 
the direct sum of the conjugate spaces spanned by $\x$ and $\vct y$ 
can be interpreted as a {\it double phase space}, where $\x$ formally plays the role of the position, 
$\vecq$, and $\vct y$ the role of its Fourier conjugate, the momentum $\vecp$. 
Then, introducing the {\it double phase space Hamiltonian},
\begin{eqnarray}
I\!\!H(\x,\vct y,t) & = & H(\x - \frac{1}{2}\mathbf J \vct y,t)
-H(\x + \frac{1}{2}\mathbf J \vct y,t) - \gamma ~ \x\cdot\vct y,
\label{defHH}
\end{eqnarray}
the above equation becomes
\begin{eqnarray}
\frac{\der \chi}{\der t} (\vct y,t) = -\frac{i}{\hbar} 
{\mathcal N}\int I\!\!H (\x,\vct y,t)
~\exp{ \left(\frac{i}{\hbar}(\vct y'- \vct y)\cdot \x \right)} 
~\chi(\vct y')~d\vct y'~d\x\cr 
- \frac{1}{2\hbar}~\Bigl[(\vl'\wedge\vct y)^2 + (\vl''\wedge\vct y)^2 \Bigr]
~\chi(\vct y,t). 
\label{dynchi_synth}
\end{eqnarray}
The same name has been kept for the characteristic function $\chi(\vct y,t)$, 
though strictly this should be $\chi(\vct \xi,t)=\chi(-\mathbf J \vct y,t)$. 

The SC approximation is built on the classical trajectories $(\x_\tau,{\vct y}_\tau)$ in the double phase space, 
with initial conditions on a Lagrangian initial surface and driven by the double Hamiltonian (\ref{defHH}) 
through Hamilton's equations: 
\begin{eqnarray} 
\dot{\vct y}_\tau  =  -\frac{\der I\!\!H}{\der \x}(\x_\tau,\vct y_\tau,\tau) ,  
\dot{\x}_\tau  =  \frac{\der I\!\!H}{\der \vct y}(\x_\tau,\vct y_\tau,\tau). 
\label{Hamilton} 
\end{eqnarray} 
In the case where $\vct y=\vct \xi=0$, the second equation above reduces to 
\be 
\dot{\x}_\tau  = \J\>\frac{\der H}{\der \x}(\x_\tau) - \gamma~ \x_\tau,
\label{centreq}
\ee 
so that the classical motion in the $\x$-plane is dissipative if $\gamma \neq 0$.
On the other hand, the evolution of chords, which are small enough to warrant 
the linearization of Hamilton's equations in the neighbourhood of $\x_\tau$, 
is driven by 
\be 
\dot{\vct y}_\tau = \left(2 \mathbf H({\x}_\tau) \mathbf J + \gamma\right) \vct y_\tau, 
\label{lindouble} 
\ee 
where $\mathbf H({\x}_\tau)$ is the Hessian matrix of $H(\x_\tau)$. In other words, small chord trajecories, ${\vct \xi}_\tau = -\J {\vct y}_\tau$, evolve according to the {\it monodromy matrix} 
for the Hamiltonian motion neighbouring the trajectory, ${ \x}_\tau$, in single phase space, 
but with a boost in the case of dissipation. Then there exists
a dissipative monodromy matrix $\M({\x}_\tau)$, 
such that ${\vct y}_\tau = \M({\x}_\tau) {\vct y}_0$
for each centre trajectory, along which is defined a quadratic Hamiltonian, 
$\x \cdot \mathbf H({\x}_\tau)~\x$. In this way, it is the time dependence through ${\x}_\tau$ 
that takes partial account of the nonlinearity of the original system.

The Markovian evolution of any chord function may be obtained by superposing linearly
the solutions of the above Lindblad equation, while substituting the density operator
by the reflection operators, $\opR_\x(t)$ in \ref{chi-R}. In other words,
one decomposes the evolution of the chord function in terms of the evolving chord symbol 
of the reflection operators, which are identified as mixed {\it centre-chord propagators} 
\cite{AlmBro06}. For unitary evolution in the limit of short chords, this reduces to
\begin{equation}
 2^N{\tilde{R}_{\x}}^0(\vct y,t) =  \exp [i\hbar^{-1}\x(t)\cdot \vct y] ~~
{\rm or}~~
2^N{\tilde{R}_{\x}}^0(\Vxi,t) =  \exp [i\hbar^{-1}\x(t)\wedge \Vxi],
\label{smallchord-R}
\end{equation}
where $\x(t)=\x_{(\tau=t)}$ is the solution of \ref{centreq} for the final time. 
Thus the propagator is just the chord symbol for a reflection operator with an evolved centre. 
The general SC Markovian form of the centre-chord propagator was obtained in \cite{AlmBro11}. 
In the SC approximation, decoherence quenches the contribution of long chords in \ref{smallchord-R}: 
\begin{equation}
 \tilde{R}_\x(\vct y,t)\approx {\tilde{R}_\x}^0(\vct y,t)
\exp\Big[\frac{-1}{2\hbar} D\{{\vct y},t;\x\}\Big]
~~{\rm or}~~ 
\tilde{R}_\x(\Vxi,t) \approx {\tilde{R}_\x}^0(\Vxi,t)
\exp\Big[\frac{-1}{2\hbar} D\{{\Vxi},t; \x\}\Big],
\label{Markovshort} 
\end{equation}
where the {\it decoherence functional} is just
\begin{equation}
 D\{{\vct y},t; \x\} = \int_0^t ~\Bigl[(\vl'\wedge{\vct y}_\tau)^2 + 
(\vl''\wedge{\vct y}_\tau)^2 \Bigr] ~d\tau
= \int_0^t ~\Bigl[(\vl'\cdot{\Vxi}_\tau)^2 + 
(\vl''\cdot{\Vxi}_\tau)^2 \Bigr] ~d\tau 
= D\{{\Vxi},t;\x\}.
\label{decfun}
\end{equation}
The limitation to short chords now allows us to consider that ${\Vxi}_\tau$ 
is itself a linear function of ${\Vxi}_0=\Vxi$,
so that \ref{decfun} is the implicit expression for a quadratic form in its components:
\be
D\{{\Vxi},t; \x\} = \Vxi \cdot {\mathbf \Phi}(t; \x)~ \Vxi,
\ee 
where the explicit form for the positive symmetric {\it decoherence matrix} is
\be
{\mathbf \Phi}(t;\x) \equiv {\int_0}^{t'} dt'~ 
{\M}({\x}_{t'-t})^\dagger(\vl'{\vl'}^\dagger + \vl''{\vl''}^\dagger){\M}({\x}_{t'-t}).
\label{decmat}
\ee
Notwithstanding the functional relation of the decoherence to the full trajectory, ${\x}_\tau$,
this in its turn is uniquely specified by its final value, $\x$.
One should recall that in the general case of several linear Lindblad operators, 
${\vl'}_k$ and ${\vl''}_k$, one merely sums each of their contributions to the exponent. 

Thus, the unitarily evolved reflection is modulated by a Gaussian.
In the simple quadratic case \cite{BroAlm04}, the monodromy matrix is independent of $\x$
and so is the decoherence matrix, ${\mathbf \Phi}(t)$: Then it is not only the chord symbol of the
reflection operator, but also any chord function is modulated by a global Gaussian, while 
the Wigner function evolves as a convolution with a widening Gaussian. There is then a universal
{\it threshold time}, $t_p$, when the Wigner function becomes positive for any initial pure state, just as $\det {\mathbf \Phi} = 1/4$. In the simplest case where the Hamiltonian motion is switched off,
we have $t_p \det (\vl'{\vl'}^\dagger + \vl''{\vl''}^\dagger) = 1/4$. Generally, 
the effect of a quadratic Hamiltonian reduces the time for positivity as is discussed in \cite{BroAlm04}:
This is a fast process in the scale of the corresponding motion of the closed system.

Inserting the Markovian centre-chord propagator in \ref{chi-R}
leads to a simple expression for the short chord limit as a superposition of atenuated reflections,
\be
\chi(\Vxi, t) = \frac{1}{(2\pi\hbar)^N}\int d\x ~W(\x)~ \exp [i\hbar^{-1}\x(t)\wedge \Vxi]
\exp\Big[\frac{-1}{2\hbar} D\{{\Vxi},t; \x\}\Big],
\label{chimarkov}
\ee
in which we will further approximate the original Wigner function by $W_c(\x)$, the classical Liouville
distribution.

\section{Wave function correlations}

Given a pure state, $\hat\rho=|\psi\rangle\langle \psi|$, a convenient definition 
of {\it local wavefunction correlations} (LWC) is
\begin{equation}
{\mathbf C}_{\Delta}(\Vxi_{\vecq}, {\mathbf Q}) \equiv  
\int \frac{d\vecq}{(\sqrt{2\pi}\Delta)^N}~ e^{-\frac{(\vecq-{\mathbf Q})^2}{2{\Delta}^2}} 
\langle \vecq - \frac{\Vxi_{\vecq}}{2}|\hat\rho|\vecq + \frac{\Vxi_{\vecq}}{2}\rangle,
\label{wavecor}
\end{equation}
where the averaging window is chosen to be SC small, that is,
\be
\Delta \rightarrow 0 ~{\rm ,~but} ~~~ \frac{\hbar}{\Delta} \rightarrow 0 ~~~~{\rm as}~~~ \hbar \rightarrow 0,
\ee
which allows for many oscillations of the wave function in the average.
An important case is where $\Delta$ is  of the order of $\hbar^{1/2}$, 
the width of a coherent state.
Though originally, a pure state was assumed, this will here be relaxed 
in admiting that decoherence may mix the state in time.
Thus, strictly speaking, there is no longer a wavefunction, so that we really study
the evolution of elements of the density matrix in position representation and their correlations, 
but there is no need to change the nomenclature. It is understood that the true correlation
should be divided by the normalization factor ${\mathbf C}_{\Delta}(0, {\mathbf Q})$,
that is, the local average of the wave intensity. Only then is \ref{wavecor}
equivalent to Berry's definition \cite{Berry77b}, but it is more convenient to
work with the formula prior to normalization.

The full wavefunction correlations, with $\Delta \rightarrow \infty$, 
may be identified with the expectations of the translation operator \ref{transop}, 
such that the translating chord has zero momentum,
\be
{\mathbf C}_{\infty}(\Vxi_{\vecq}) = {\rm tr}~\hat\rho~\hat{T}_{(\Vxi_{\vecq}, 0)}.
\ee
It is then expedient to define {\it local translation operators}
\begin{equation}
\hat{\bf t}_{\Vxi_{\vecq}, \mathbf Q, \Delta}  \equiv 
\int \frac{d\vecq}{(\sqrt{2\pi}\Delta)^N} \left| \vecq + \frac{\Vxi_{\vecq}}{2} \right\rangle 
\left\langle \vecq - \frac{\Vxi_{\vecq}}{2} \right|
                e^{-\frac{(\vecq-{\mathbf Q})^2}{2{\Delta}^2}}   ,
\label{localtrans}								
\end{equation}
in a similar fashion to the full translation operators \ref{transop}, 
but only transitions in position are considered and these have a finite range, $\Delta$, around $\mathbf Q$. Then one reinterprets the LWC as the expectation of a local translation:
\be 
{\mathbf C}_{\Delta}(\Vxi_{\vecq}, {\mathbf Q})  
= \left \langle \hat{\bf t}_{\Vxi_{\vecq}, \mathbf Q, \Delta} \right\rangle
= {\rm tr}~\hat\rho~\hat{\bf t}_{\Vxi_{\vecq}, \mathbf Q, \Delta}  ~.
\label{wavecor1}
\ee

If one now expresses the density operator as a superposition of reflexions \ref{chi-R},
the LWC also become a superposition of the {\it elementary correlations}, 
which are just the Weyl representation of the local translation operator,
\begin{eqnarray}
{\bf t}_{\Vxi_{\vecq}, \mathbf Q, \Delta}(\x)
&& \equiv  {\rm tr}~ 2^N\hat{R}_\x~\hat{\bf t}_{\Vxi_{\vecq}, \mathbf Q, \Delta}\nonumber  \\
&& = \frac{1}{(\sqrt{2\pi}\Delta)^N}
\exp\left\{-\frac{i}{\hbar}\vecp\cdot {\mathbf \Vxi_{\vecq}} ~
-\frac{1}{2{\Delta}^2}({\mathbf Q}-\vecq)^2\right\}, 
\label{elcor}
\end{eqnarray}
which cut off the plane waves \ref{planewave}, representing full translations by 
$\Vxi=(\Vxi_\vecq, 0)$,  outside of the Gaussian window of positions.

The spectrum of a single elementary correlation, that is, its Fourier transform with respect to
$\Vxi_{\vecq}$ is just a $\delta$-function,
\be
{\$}_{\vecp',\mathbf Q,\Delta}(\x) =\frac{1}{(\sqrt{2\pi}\Delta)^N}  \delta({\vecp}' - \vecp)~ 
e^{-\frac{1}{2{\Delta}^2}({\mathbf Q}-\vecq)^2},
\label{elspec}
\ee
where only the amplitude depends on $\mathbf Q$.
It is important to note that the local translation operator is not an observable.
However,
\be
{\hat O}_+ = (\hat{\bf t}_{\Vxi_{\vecq}, \mathbf Q, \Delta}
+\hat{\bf t}_{-\Vxi_{\vecq}, \mathbf Q, \Delta}) ~~~
{\rm and}~~~ 
{\hat O}_- = i(\hat{\bf t}_{\Vxi_{\vecq}, \mathbf Q, \Delta}-\hat{\bf t}_{-\Vxi_{\vecq}, \mathbf Q, \Delta})
\label{lobservable}
\ee
are simple mechanical observables, with real Weyl symbols that equal twice the real and the imaginary
parts of \ref{elcor}. The labels of the translation have here been ommited, but in general there will be a
doubling of the correlations that are discussed in this paper. For instance, the elementary correlations
then give rise to a pair of terms such as \ref{elcor}, so that the full spectrum has peaks at $\pm \vecp$.

The chord symbol for the local translation is the Fourier transform of \ref{elcor}:
\be
{\tilde {\bf t}}_{\Vxi_{\vecq}, \mathbf Q, \Delta}(\vct \eta) =
\delta{( {\vct \eta}_\vecq - \Vxi_\vecq)}~~
\exp\left\{\frac{i}{\hbar}{\vct \eta}_{\vecp}\cdot {\mathbf Q} ~
-\frac{{\Delta}^2}{2{\hbar}^2}{\vct \eta}_{\vecp}^2 \right\},
\ee
so that the chord representation of \ref{wavecor1},
\be
 {\mathbf C}_{\Delta}(\Vxi_{\vecq}, {\mathbf Q}) 
= \int {d\vct \eta}~ \chi(\vct \eta)~ {\tilde {\bf t}}_{\Vxi_{\vecq}, \mathbf Q, \Delta}(-\vct \eta) 
= \int d{\Vxi}_{\vecp} ~ \chi({\Vxi}_{\vecq},{\Vxi}_{\vecp})~
\exp \left\{\frac{i}{\hbar}{\Vxi}_{\vecp}\cdot {\mathbf Q} ~
-\frac{{\Delta}^2}{2{\hbar}^2}{\Vxi}_{\vecp}^2\right\},
\label{wavecor2}
\ee
retrieves the exact expression for the LWC as a smoothed Fourier transform of the chord function, 
derived in \cite{ZamZapOA}. 
It should be recalled that the identification of the chord function 
in an arbitrarily small neighbourhood of the origin
with the Fourier transform of the Liouville distribution, $W_c(\x)$, determines all moments 
to be entirely classical. By extending this region to a {\it Planck volume}, that is, 
to the classically small {\it Planck volume}, $(2\pi \hbar)^N$ of a coherent state for $N$ degrees of freedom, 
one has access to purely quantum features, in spite of the classical basis 
of our approximate chord function.   
Choosing the smoothing parameter as $\Delta\approx {\hbar}^{1/2}$, i.e. the
width of a coherent state, the Gaussian integration window for positions \ref{wavecor1} 
has a volume $\hbar^{N/2}$. Within this classically small volume, a SC wave function, 
such as a highly excited eigenstate, necessarily involves an average over many oscillations
of wave length $\hbar$. Then the same Gaussian width is obtained for the smoothing 
of the chord function in \ref{wavecor2}, so that the LWC
only require the information of the chord function within a Planck volume surrounding the origin.

Note that the simple expression for the chord function of a state
that is translated in phase space by a chord $\vct \eta$,
\be
\chi_{\mathbf \eta}(\Vxi) = \chi(\Vxi)~ e^{\frac{i}{\hbar}\Vxi\wedge \vct \eta},
\label{transchi}
\ee
leads to the reading of \ref{wavecor2} as a mere Gaussian smoothed {\it projection}
of the chord function that is translated by $\vct \eta = ({\mathbf Q}, 0)$. 
 
Alternatively, the definition of the quenched chord function,
\be
\chi_\Delta(\Vxi) \equiv \chi(\Vxi) ~\exp \left\{-\frac{{\Delta}^2}{2{\hbar}^2}{\Vxi}_{\vecp}^2\right\}.
\label{chidelta}
\ee
allows the valid interpretation of $ {\mathbf C}_{\Delta}(\Vxi_{\vecq}, {\mathbf Q})$
as its inverse chord transform,
$\langle \mathbf{Q} - \frac{\Vxi_{\vecq}}{2}|\hat\rho_\Delta|\mathbf{Q} + \frac{\Vxi_{\vecq}}{2}\rangle$.
Comparison with \ref{chimarkov} shows that this is a legitimate mixed chord function, which could have
resulted from a Markovian evolution driven by a quadratic Hamiltonian. In other words, the matrix element,
$\langle \mathbf{Q} - \frac{\Vxi_{\vecq}}{2}|\hat\rho|\mathbf{Q} + \frac{\Vxi_{\vecq}}{2}\rangle$,
resulting from the inverse transform of $\chi(\Vxi)$, is locally averaged in the correlation window 
in the same way as would be achieved by the nonunitary evolution $\hat\rho \mapsto \hat\rho_\Delta$ 
of the density operator, represented by \ref{chidelta}, as discussed in the previous section. 
Thus the effect of decoherence is neatly included into the SC analysis of the LWC in the following section.

A simple example of the LWC arises for a coherent state, labeled by the vector
$\vct \eta=(\vct \eta_p,\vct \eta_q)$,
\begin{eqnarray}
\langle \vecq|{\vct \eta}\rangle= \Big(\frac{\omega}{\pi \hbar}\Big)^{N/4}\exp\Big(-\frac{\omega}
{2\hbar}(\vecq-\vct \eta_q)^2+\frac{i}{\hbar} \vct\eta_p \cdot (\vecq-\frac{\vct\eta_q}{2})\Big).
\label{costate}
\end{eqnarray}
Its chord function for $\omega=1$ is just
\begin{eqnarray}
 \chi_{\vct \eta}(\Vxi) =  \frac{1}{(2\pi \hbar)^N}
\exp\left[\frac{i}{\hbar} {\vct \eta} \wedge \Vxi - \frac{1}{4\hbar}\Vxi^2 \right]  ,
\end{eqnarray}
which is of the same form as \ref{Markovshort}. Thus, the chord function for a pure coherent state
equals the chord symbol for a reflection that has undergone an appropriate decoherence.
The LWC is then
\be 
{\mathbf C}_{\Delta}(\Vxi_{\vecq}, {\mathbf Q}; \vct \eta) =
\frac{1}{(\sqrt{2\pi}\Delta)^N}
\exp\left\{-\frac{i}{\hbar}\vct \eta_p \cdot {\Vxi_q} ~
-\frac{1}{2\hbar} {\Vxi_\vecq}^2
-\frac{1}{2({\Delta}^2+\hbar)}({\mathbf Q}-\vct \eta_q)^2\right\},
\label{corcoherent}
\ee
with the spectrum
\be
{\$}_{\vecp',\mathbf Q,\Delta}(\vct \eta) =
\frac{1}{(\sqrt{2\pi}\Delta)^N}  
\exp\left[\frac{1}{2\hbar}({\vecp}' - \vct \eta_p)^2 -
\frac{1}{2(\Delta^2+\hbar)}({\mathbf Q}-\vct \eta_q)^2 \right]~.
\label{specoherent}
\ee
Thus, there is a single spectral peak that is centred on the classical momentum with a width of $\sqrt \hbar$.

\section{Semiclassical correlations}

In the case where the SC state corresponds to a torus specified by action-angle variables, 
with constant $\vct I$ so that $\x = \x(\vct I, \vct \theta) $, the short chord approximation 
corresponding to \ref{rho_c} is 
\begin{equation}
\chi_{\vct I}(\Vxi) = \frac{1}{(2\pi\hbar)^N}\int \frac{d{\vct \theta}}{(2\pi)^N} ~ 
\exp \left\{\frac{i}\hbar \,\x(\vct I, \vct \theta) \wedge \Vxi\right\}.
\label{intsc}
\end{equation}
Inserting this into \ref{wavecor2} leads to
\begin{eqnarray}
&& {\mathbf C}_{\Delta}(\Vxi_{\vecq}, {\mathbf Q}) = 
\int \frac{d{\vct \theta}}{(2\pi)^N} ~ 
\exp \left\{\frac{i}\hbar \,\vecp(\vct I, \vct \theta) \cdot \Vxi_{\vecq}
- \frac{1}{2{\Delta}^2}({\mathbf Q}-\vecq(\vct I, \vct \theta))^2\right\}\nonumber  \\
&& = \sum_j \int d\vecq~ \left\|\det \frac{\der \vct \theta_j}{\der\vecq}\right\| 
\exp \left\{\frac{i}\hbar \,\vecp_j(\vecq) \cdot \Vxi_{\vecq}
-\frac{1}{2{\Delta}^2}({\mathbf Q}-\vecq)^2\right\}~,
\label{wavecor3}
\end{eqnarray}
where now the $N$ angles become multivalued functions of the positions, given the actions $\vct I$.
If $\Delta$ is small within a classical scale, we may fix
\be
\vecp_j(\vecq) \approx \vecp_j(\vct Q) ~~ {\rm and} ~~ 
\left\|\det \frac{\der \vct \theta_j}{\der\vecq}\right\| 
\approx \left\|\det \frac{\der \vct \theta_j}{\der\vct Q}\right\|,
\ee
which leads to Berry's approximation \cite{Berry77b} of the LWC as
\be
{\mathbf C}_{\Delta}(\Vxi_{\vecq}, {\mathbf Q}) \approx 
\sum_j \left\|\det \frac{\der \vct \theta_j}{\der\vct Q}\right\|
\exp \left\{\frac{i}\hbar \,\vecp_j(\vct Q) \cdot \Vxi_{\vecq}\right\},
\label{wavecor4}
\ee
within a normalization.
It should be recalled that
\be
\left\|\det \frac{\der \vct \theta_j}{\der\vct Q}\right\|
= \left\|\det \frac{{\der}^2 S_j}{\der\vct Q \der \vct I}\right\|
= \left\|\det \frac{\der {\vct I}(\vct Q, \vecp_j(\vct Q)}{\der\vecp}\right\|^{-1}
\label{waveamp}
\ee
is just the square of the amplitude for the j'th term of the SC wavefunction, defined by the
j'th branch of its action function, $S_j(\vecq, \vct I)$. These branches 
are separated by {\it caustics} where \ref{waveamp} is singular. One should further note 
that the absence of time in these formulae in no way precludes the treatment 
of an unitary evolution. Then the SC approximations merely relie on a classically evolving
Lagrangean surface, so there is an implicit time dependence of the action and angle coordinates, 
$\x(\vct I, \vct \theta, t)$, and likewise the branches of the action function, 
$S_j(\vecq, \vct I, t)$, will be separated by moving caustics.

An improvement on the above approximation results from the inclusion 
of the second term in the expansion,
\be \vecp_j(\vct Q) + \frac{\der \vecp_j}{\der\vct Q} (\vecq - \vct Q)
= \vecp_j(\vct Q) + \frac{{\der}^2 S_j}{\der\vct Q\der \vct Q} (\vecq - \vct Q),
\label{linp}
\ee
within the exponent of the integral in \ref{wavecor2}, so that 
\begin{eqnarray}
{\mathbf C}_{\Delta}(\Vxi_{\vecq}, {\mathbf Q}) \approx &&
\sum_j \exp \left\{\frac{i}{\hbar} \,\vecp_j(\vct Q) \cdot \Vxi_{\vecq}\right\}\nonumber  \\
&&\int d\vecq~ \left\|\det \frac{\der \vct \theta_j}{\der\vecq}\right\| 
\exp \left\{\frac{i}{\hbar} \,\Vxi_{\vecq} \cdot 
\frac{{\der}^2 S_j}{\der\vct Q\der \vct Q} ({\mathbf Q}-\vecq)
-\frac{1}{2{\Delta}^2}({\mathbf Q}-\vecq)^2\right\}, 
\label{wavecor5} 
\end{eqnarray}
which then integrates to
\be
{\mathbf C}_{\Delta}(\Vxi_{\vecq}, {\mathbf Q}) \approx 
\sum_j  \left\|\det \frac{\der \vct \theta_j}{\der\vct Q}\right\|
\exp \left\{\frac{i}\hbar \,\vecp_j(\vct Q) \cdot \Vxi_{\vecq} 
- \frac{{\Delta}^2}{2{\hbar}^2} \Vxi_\vecq \cdot 
\left(\frac{{\der}^2 S_j}{\der\vct Q\der \vct Q}\right)^2 \Vxi_\vecq\right\}~.
\label{wavecor6}
\ee
Thus one arrives at a discrete superposition of {\it wavelets} in the correlation length, 
which unlike the elementary correlations \ref{elcor} and the simple approximation \ref{wavecor4}, 
have a broadened frequency spectrum,
\be
 \$_{{\vecp}', {\mathbf Q}, \Delta} \approx
\sum_j \left\|\det \frac{\der \vct \theta_j}{\der\vct Q}\right\|
\exp \left\{-\frac{1}{2{\Delta}^2} (\vecp'-\vecp_j(\vct Q)) \cdot 
\left(\frac{{\der}^2 S_j}{\der\vct Q\der \vct Q}\right)^{-2} (\vecp'-\vecp_j(\vct Q))\right\}~, 
\label{intspec}
\ee
though the widths go to zero in the case of a {\it flat} Lagrangean surface, that is, 
$\frac{\der \vecp_j}{\der\vecq}=0$, such as in a billiard. Otherwise, the spectrum
still has narrow peaks in comparison to their separation, $\vecp_j(\vct Q)- \vecp_{j'}(\vct Q)$, 
if one chooses ${\Delta}^2\approx \hbar$ as discussed in the Appendix.
It is only when the averaging width $\Delta$ is chosen of the order of these separations
that the individual lines will no longer be detectable.

Before proceeding further, it is worthwhile to introduce the tangent vectors 
to the Lagrangian surface at the point $\x_j = (\vct Q, \vecp_j(\vct Q))$, that is,
\be
\Vxi^j =({\Vxi^j}_\vecq, {\Vxi^j}_\vecp) 
\equiv \left({\Vxi}_\vecq, \frac{\der \vecp_j}{\der\vct Q}{\Vxi}_\vecq\right),
\label{tanchord}
\ee
for any arbitrary chord component $\Vxi_\vecq$. This allows for the compactification
of the approximation \ref{wavecor6} as
\be
{\mathbf C}_{\Delta}(\Vxi_{\vecq}, {\mathbf Q}) \approx 
\sum_j  \left\|\det \frac{\der \vct \theta_j}{\der\vct Q}\right\|
\exp \left\{\frac{i}\hbar \,\vecp_j(\vct Q) \cdot \Vxi_{\vecq} 
- \frac{{\Delta}^2}{2{\hbar}^2}  ({\Vxi^j}_\vecp)^2\right\}~,
\label{wavecor7}
\ee
which will be convenient for the analysis of Markovian evolution.

The effect of decoherence within a scenario dominated by short chords is now incorporated
through the SC approximation \ref{Markovshort}, as specified by the quadratic form in terms of 
the decoherence matrix \ref{decmat}, within the integral in \ref{wavecor6}. 
The same expansion of the phase as in the unitary case around the classically relevant points,
$\x_j =(\mathbf Q, \vecp_j(\mathbf Q))$, while the decoherence matrix for each of these is fixed at  
\be
\mathbf \Phi(t, \x_j) \equiv {\mathbf \Phi}^j,
\ee
then results in the inclusion of a second Gaussian term in the integral. In other words,
the modified chord function \ref{chidelta} merely picks up a true local decoherence factor on top
of the correlation smoothing:
\be
\chi_\Delta(\Vxi) \mapsto \chi(\Vxi) ~\exp \left\{-\frac{{\Delta}^2}{2{\hbar}^2}{\Vxi}_{\vecp}^2\right\}
\exp \left\{-\frac{1}{2\hbar}\Vxi \cdot {\mathbf \Phi}^j ~ \Vxi\right\}.
\label{chidelta'}
\ee

The result can then be expressed in terms of the tangent vectors defined by \ref{tanchord},
so that the SC approximation for the Markovian evolution of the LWC takes the form 
of a simple generalization of \ref{wavecor7}:
\be
{\mathbf C}_{\Delta}(\Vxi_{\vecq}, {\mathbf Q}) \approx
\sum_j  \left\|\det \frac{\der \vct \theta_j}{\der\vct Q}\right\|
\exp \left\{\frac{i}\hbar \,\vecp_j(\vct Q) \cdot \Vxi_{\vecq} 
- \frac{1}{2\hbar} \Vxi^j \cdot {\mathbf \Phi}^j \Vxi^j
- \frac{{\Delta}^2}{2{\hbar}^2}  ({\Vxi^j}_\vecp)^2\right\}~. 
\label{wavecor8}
\ee

The broadening of the spectral correlations depends on the full quadratic form, 
$\Vxi^j \cdot {\mathbf \Phi}^j \Vxi^j$, even though the explicit dependence 
on the correlation length itself, $\Vxi_\vecq$, arises indirectly through \ref{tanchord}. 
To find the explicit dependence, 
one makes use of the {\it symplectic invariance} of the entire centre-chord formalism 
\cite{BroAlm04}, that is, our results are equally valid
after symplectic transformations are performed in phase space, which correspond to 
{\it metaplectic} quantum transformations\cite{Littlejohn86}: 
If the classical transformation is 
$\x \mapsto \x'=\mathbf C~ \x$, then the transformed decoherence matrix is
\be
\mathbf \Phi'(t, \x') \equiv {\mathbf C}^\dagger~ \mathbf \Phi(t, \x)~ \mathbf C.
\ee 
Let us then define the symplectic shear transformations, ${\mathbf C}_j$, that are taylored
to bring each tangent plane at $\x_j$ to the horizontal:
\begin{eqnarray}
&&\vecp'-\vecp_j(\vct Q) = \vecp - \vecp_j(\vct Q) - \frac{\der \vecp_j}{\der\vct Q}~ (\vecq - \vct Q) \nonumber \\
&&\vecq'-\vct Q = \vecq - \vct Q ~.
\end{eqnarray}
But the chords transform in the same way as $\x- \x_j$, so that the shearing takes $\Vxi_j \mapsto\Vxi'_j = (\Vxi_\vecq, 0)$, 
which brings the decoherence matrix in \ref{wavecor8} to the simple form
\be
\Vxi^j \cdot {\mathbf \Phi}^j \Vxi^j = {\Vxi'}^j \cdot {\mathbf \Phi'}^j {\Vxi'}^j
= \Vxi_q \cdot {\mathbf \Phi'}^j_{\vecq\vecq} \Vxi_q,
\ee
with 
\be
{\mathbf \Phi'}^j = \mathbf \Phi'(t, {\x'}_j) ={\mathbf C_j}^\dagger~ \mathbf \Phi^j~ \mathbf C_j.
\ee
Thus, one arrives at the alternative form of the Markovian evolution of the LWC,
\begin{eqnarray}
{\mathbf C}_{\Delta}(\Vxi_{\vecq}, {\mathbf Q}) \approx &&
\sum_j  \left\|\det \frac{\der \vct \theta_j}{\der\vct Q}\right\| \nonumber \\
&&\exp \left\{\frac{i}\hbar \,\vecp_j(\vct Q) \cdot \Vxi_{\vecq} 
- \frac{1}{2\hbar} \Vxi_\vecq \cdot {\mathbf \Phi'}^j_{\vecq\vecq} \Vxi_\vecq
- \frac{{\Delta}^2}{2{\hbar}^2}  \Vxi_\vecq \cdot 
\left(\frac{{\der}^2 S_j}{\der\vct Q\der \vct Q}\right)^2 \Vxi_\vecq\right\}~. 
\label{wavecor9}
\end{eqnarray}
which depends explicitly on the correlation parameter, $\Vxi_\vecq$. 

The Fourier transform in relation to $\Vxi_\vecq$ then leads to the generalization of the spectrum of the LWC for a pure state,
given by \ref{intspec}, as
\begin{eqnarray}
\$_{{\vecp}', {\mathbf Q}, \Delta} \approx &&
\sum_j \left\|\det \frac{\der \vct \theta_j}{\der\vct Q}\right\| \nonumber \\
&&\exp \left\{-\frac{1}{2\hbar} (\vecp'-\vecp_j(\vct Q)) \cdot 
\left[{\mathbf \Phi'}^j_{\vecq\vecq} + 
\frac{\Delta^2}{\hbar}\left(\frac{{\der}^2 S_j}{\der\vct Q\der \vct Q}\right)^{2}\right]^{-1} 
(\vecp'-\vecp_j(\vct Q))\right\}~. 
\label{intspec2}
\end{eqnarray}
In the case of a single degree of freedom, the matrices, ${\mathbf \Phi'}^j_{\vecq\vecq}$, reduce to a scalar.
So that, after the short short time of the positivity threshold, $t_p$, discussed in section 3, decoherence dominates and the width of the peak is basically $\sqrt{\hbar~ {\mathbf \Phi'}^j_{\vecq\vecq}}$.

\section{Discussion}

The Markovian approximation for general open evolution of quantum systems only holds 
in the limit of weak coupling to the environment. Even so, the qualitative features
of decoherence and quantum dissipation are already integral features of the Markovian
scenario. The introduction of corrections, related to the memory kept by the environment
of the system's past motion, denies the formulation of a simple master equation 
that is purely differential, thus impairing a fully general description.
We conjecture that the expectation of the local translation operators, that is, the LWC
belong to the robust set of features that are at least qualitatively captured by
the Markovian framework. 

The combination of the Markovian, the semiclassical and the short chord aprroximations wraps 
the effect of decoherence into a single quadratic form, $\Vxi^j \cdot {\mathbf \Phi}^j \Vxi^j$.
This broadens spectral lines in the Fourier transform of the LWC, 
as the elements of each matrix, ${\mathbf \Phi}^j$, 
grow with decoherence. Another important ingredient is \ref{tanchord}, determining
that this quadratic form is an increasing function of the  
of the slope of the Lagrangian surface at each point $\x_j=(\vct Q, \vecp_j(\vct Q))$,
for any given correlation vector, $\Vxi_\vecq$. Thus, the points $\x_j$ 
that are close to a caustic (where $\frac{\der \vecp_j}{\der\vct Q}\rightarrow \infty$) 
contribute wider lines to the spectrum. But it is just at a caustic that the Lagrangian branches, 
$\vecp_j(\vct Q)$, coalesce, so it is in the neighbourhood of a caustic 
that the width of the spectral peaks first become wider than their separation,
as decoherence increases. 
\begin{figure}[htb!]
\centering
\includegraphics[height=8cm]{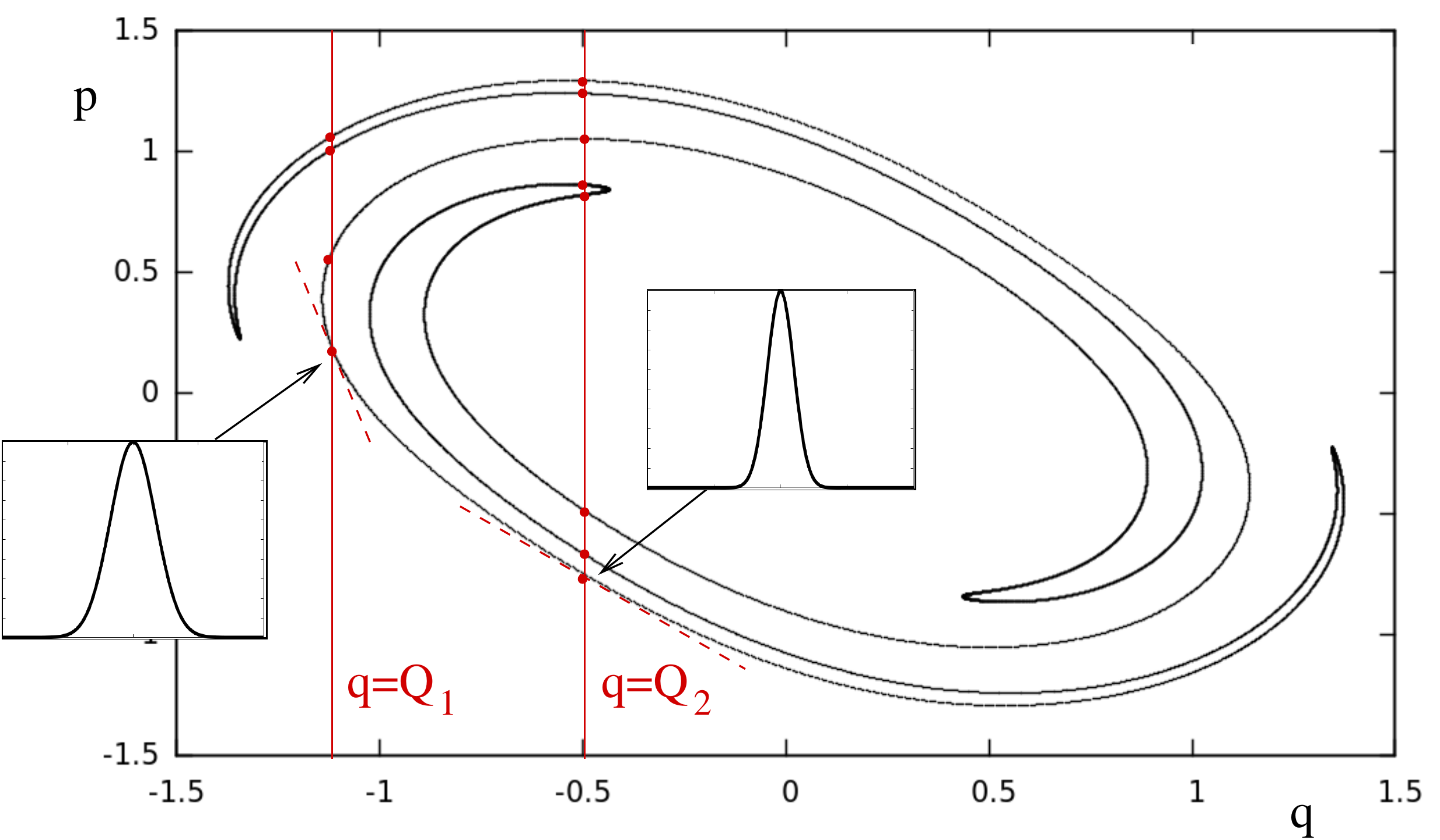}
\caption{A simple initially convex curve typically develops many sheets, $\vecp_j(\vct q)$. 
Then the LWC spectrum for the SC state corresponding to the above curve has 4 peaks at ${\vct Q}_1$
and 8 peaks at ${\vct Q}_2$. The distance between the peaks is just that of the corresponding classical momenta at each point. The width of each peak increases with the slope of the curve. This is maximal near
a caustic, where a pair of branches coalesce. All peaks broaden with decoherence, which ultimately destroys their resolution.}
\label{Fig1}
\end{figure}

The simplest case is that of an initial eigenstate of the driving Hamiltonian,
or an eigenstate of observables which commute with it.
Then the Lagrangian surface is fixed and all the spectral broadening
depends only on the growth of the decoherence matrices. The last spectral lines
to disapear with decoherence are those which have the centre of the correlation window, $\vct Q$, 
far from the fixed caustics. In contrast, the evolution of a general SC state will correspond 
to a Lagrangian surface that wraps around itself in complicated ways, develping 'whorls' and 'tendrils' 
as discussed in \cite{BerBal79}. Thus, a convex closed curve can evolve into something
as shown in Fig 1. The number of branches increases and more peaks are produced in the LWC spectrum:
They are closer together and they are broadened by the viccinity of more caustics
joining these new branches. In this way, the purely classical bending of the Lagrangian
surface is partly responsible for the loss of quantum wavyness.

So far we have always assumed that the initial state is a generalized WKB state,
corresponding to a single Lagrangian surface. Of course one can also allow superpositions
of such states. These lead to interferences in the complete Wigner function,
which become large structures in the chord function. The first effect of decoherence
is to quench the amplitude of all large chords and thus to erase large structures
in the chord space, as opposed to the long survival of the spectral peaks of the LWC, 
which depend only on short chords. Hence, the LWC for a superposition
is no different from th the at of a mixture of the same states. 
Indeed, this is yet another indication of the robustness of the LWC:
The LWC for the mixture of a few states preserves the independent spectral peaks
of each component state, so the mixture has more of them and their average distance diminishes. 
Nonetheless, the peaks in the components of the mixture can still be resolved, 
as long as the peak widths are smaller than all the separations.

An initial coherent state also deviates from our previous assumptions. 
Indeed, the LWC spectrum for these states \ref{specoherent} has a single peak
with a width $\sqrt\hbar$. Nonetheless, the spectrum of the expectation for the
symmetrizations of the local translation operator \ref{lobservable} doubles the number of peaks,
so that the peaks for a coherent state centred at the phase space point, $\vct \eta$,
lie at $\vecp' = \pm \vct \eta_\vecp$. Thus, as far as concerns the detection of wavyness,  
a coherent state closer to the origin is more classical than one far from the origin.
\footnote{This is foreign to the usual phase space view of quantum mechanics and contradicts
the quantum optics picture, in which the far lying coherent states have more photons
and so their field is considered to be more classical.}

The nonlinear evolution of such a state quickly drives it
into a thin surface or line, in the case of a single degree of freedom,
with an oscillating Wigner function, as described by \cite{TomHel91,TomHel93,MNVT}. 
Decoherence will quench this long chord structure, within a typical threshold time, $t_p$,
for positivity of the Wigner function discussed in section 3. This may even precede
the formation of the interferences in an unitary evolution.
But the intrinsic wavyness portrayed by LWC should survive much longer
and we conjecture that it can be calculated by the classical short chord
approximation to the chord function \ref{rho_c}, even for the evolution of 
the initial classical Gaussian distribution corresponding to the coherent state.

Special care should be taken in the adaptation of the present results to
quantum billiards. Indeed, the Wigner and the chord transforms themselves 
are affected by the boundary, which needs to be considered in computational
work on wave function correlations for billiards. 
The loss of dependence on the position, $\vct Q$,
of the averaging window, as long as this is far from the boundary, simplifies calculations.
Collisions of an initially smooth Lagrangian surface with the boundary will separate it into
disconnected branches that are no longer connected by caustics, but our previous discussion
about the relation of peak widths and their separations still holds.

All in all, one can be sure that the emergence of positivity of the Wigner function through
decoherence does not necessarily imply the absence of detectable wavyness in the 
evolving mixed state. Positivity requires that the elements of the decoherence matrices are of order $\hbar^0$,
so that the resulting mixed Wigner function convoluted with a Gaussian with a width
of order $\hbar^{1/2}$ resembles a pure Husimi function.
This occurs globally at a precise positivity time, $t_p$, if the Hamiltonian is quadratic,
and picemeal otherwise, but within a similar overall scale. 
Then each LWC contribution also has a width of order $\hbar^{1/2}$ 
and so do the corresponding spectral peaks. In contrast, the centres of these peaks
are located at $\vecp_j(\vct Q)$, i.e. at the relevant classical momenta. 
So, in the case of a single freedom, it is only when the spectral widths, 
$\sqrt{\hbar~ {\mathbf \Phi'}^j_{\vecq\vecq}} > |\vecp_j(\vct Q)-\vecp_{j'}(\vct Q)|$,
that the peaks can no longer be resolved. For this to happen, the elements 
of the decoherence matrices must have grown to be of order $\hbar^{-1}$.
Thus, one should be able to detect residual quantum wavyness,
identified as the expectation of a local translation operator, 
long after positivity of the Wigner function has been reached.

\appendix
\section{Relation between the Husimi function and the LWC}

Given a coherent state \ref{costate}
with its Gaussian Wigner function,
\begin{equation}
\label{wcoherent}
W_{\vct\eta}(\x) = \frac{1}{(\pi \hbar)^N} e^{-(\x-\vct\eta)^2/\hbar}  ,
\end{equation}
the Husimi function is the average 
\be
\rho_H(\vct\eta)= \langle\vct\eta|\rho|\vct\eta\rangle = {\rm tr}\>\rho\>|\vct\eta \rangle\langle \vct\eta|,
\label{Husimi1}
\ee
which is non-negative for all phase space points, $\vct\eta$.
Moreover, it can also be interpreted as a smoothed Wigner function, 
\be
\rho_H(\vct\eta)=\int \rm d\x\>W_{\vct\eta}(\x)\>W(\x),
\label{Husimi2}
\ee
that is, the convolution of the Wigner function with a phase space Gaussian of volume $\hbar^N$.
It follows that the Fourier transform of the Husimi function is just
\be
F(\Vxi) = \frac{1}{(2\pi \hbar)^N}~e^{-\Vxi^2/{4\hbar}}~ \chi(\Vxi) ,
\label{FTH}
\ee
so that the long chords are suppressed and only the neighbourhood of the origin with radius $\hbar^{1/2}$
is appreciable.

The subtlety is that the complete encoding of quantum information in the Husimi function is so
delicate as to disallow us to neglect the evanescent region in the Gaussian tail, for it enters
in the inverse Fourier transform for the complete Husimi function. Even so, 
the Husimi function of a pure state can be reconstructed from the knowledge of the LWC, 
even though they are also obtained by coarse graining the chord function.
To see this, consider the effect of a further smooth projection, but now onto the ${\mathbf P}$-axis
with the specific choice $\Delta = {\hbar}^{1/2}$:
\be
 \int \frac{d\Vxi_{\vecq}}{(2\pi\hbar)^N}~{\mathbf C}_{\Delta}(\Vxi_{\vecq},{\mathbf Q})~
e^{-\frac{i}{\hbar}\Vxi_{\vecq}\cdot {\mathbf P}} ~
e^{-\frac{{\Delta}^2}{2{\hbar}^2}\Vxi_{\vecq}^2}
= \int \frac{d\Vxi}{(2\pi\hbar)^N}~ e^{-\frac{i}{\hbar}\Vxi\wedge {\mathbf X}} ~
\chi(\Vxi)~ e^{-\frac{{\mathbf \Vxi}^2}{2\hbar}}
= {\rho_H(X)}.
\label{Husimi3}
\ee
Thus, we recognise the Fourier transform of \ref{FTH}, that is, the Husimi function, 
evaluated at the phase space point, ${\mathbf X}=({\mathbf Q,{\mathbf P}})$, 
so that exact knowledge of the local correlation for all positions, 
${\mathbf Q}$, also contains, in principle, complete information about the state.
In spite of smoothing the wave function to obtain the correlation, 
the fact that according to \ref{transchi} one translates the chord for each position 
then implies that the short chord information for the set of translated states 
has the full information of  the single original state.

\acknowledgements{I thank Olivier Brodier for the figure, as well as stimulating discussions. Partial financial support 
from the National Institute for Science and Technology: Quantum Information, FAPERJ and CNPq is gratefully acknowledged.}

\end{document}